\documentclass[twocolumn,superscriptaddress,nofootinbib,floatfix]{revtex4-1}
\pdfoutput=1 
\usepackage{amsmath}
\usepackage{amssymb}
\usepackage{graphicx}
\usepackage{color}
\usepackage{soul}
\usepackage{float}
\usepackage{hyperref}
\usepackage{epsfig}
\usepackage{epstopdf}
\usepackage{natbib}
\usepackage[utf8]{inputenc} % allow utf-8 input
\usepackage[T1]{fontenc}    % use 8-bit T1 fonts
\usepackage{url}            % simple URL typesetting
\usepackage{booktabs}       % professional-quality tables
\usepackage{amsfonts}       % blackboard math symbols
\usepackage{nicefrac}       % compact symbols for 1/2, etc.
\usepackage{microtype}      % microtypography
\usepackage[table]{xcolor}
\usepackage{colortbl}
\usepackage{multirow}

\definecolor{midgreen}{rgb}{0.52, 0.73, 0.4}

\newcommand{\beq}{\begin{equation}}
\newcommand{\eeq}{\end{equation}}
\newcommand{\bea}{\begin{eqnarray}}
\newcommand{\eea}{\end{eqnarray}}

\begin{document}

\title{Influence of heavy resonances in SMASH}

\author{Jordi Salinas San Martin}
\affiliation{Illinois Center for Advanced Studies of the Universe, Department of Physics, University of Illinois at Urbana-Champaign, Urbana, IL 61801, USA}
\author{Jacquelyn Noronha-Hostler}
\affiliation{Illinois Center for Advanced Studies of the Universe, Department of Physics, University of Illinois at Urbana-Champaign, Urbana, IL 61801, USA}
\author{H. Elfner}
\affiliation{GSI Helmholtzzentrum für Schwerionenforschung, Planckstr. 1, 64291 Darmstadt, Germany}
\affiliation{Institute for Theoretical Physics, Goethe University, Max-von-Laue-Strasse 1, 60438 Frankfurt am Main, Germany}
\affiliation{Frankfurt Institute for Advanced Studies, Ruth-Moufang-Strasse 1, 60438 Frankfurt am Main, Germany}
\affiliation{Helmholtz Research Academy Hesse for FAIR (HFHF), GSI Helmholtz Center,\\
Campus Frankfurt, Max-von-Laue-Strasse 12, 60438 Frankfurt am Main, Germany}
\author{J. Hammelmann}
\affiliation{Institute for Theoretical Physics, Goethe University, Max-von-Laue-Strasse 1, 60438 Frankfurt am Main, Germany}
\affiliation{Frankfurt Institute for Advanced Studies, Ruth-Moufang-Strasse 1, 60438 Frankfurt am Main, Germany}
\author{R. Hirayama}
\affiliation{Institute for Theoretical Physics, Goethe University, Max-von-Laue-Strasse 1, 60438 Frankfurt am Main, Germany}
\affiliation{Frankfurt Institute for Advanced Studies, Ruth-Moufang-Strasse 1, 60438 Frankfurt am Main, Germany}
\affiliation{Helmholtz Research Academy Hesse for FAIR (HFHF), GSI Helmholtz Center,\\
Campus Frankfurt, Max-von-Laue-Strasse 12, 60438 Frankfurt am Main, Germany}

\begin{abstract}
Recent lattice QCD results, comparing to a hadron resonance gas model, have shown the need for hundreds of particles in hadronic models. These extra particles influence both the equation of state and hadronic interactions within hadron transport models.
Here, we introduce the PDG21+ particle list, which contains
the most up-to-date database of particles and their properties.  We then convert all particles decays into 2 body decays  so that they are compatible with SMASH in order to produce a more consistent
description of a heavy-ion collision.
\end{abstract}

\maketitle

\section{Introduction}
In the 1960's Rolf Hagedorn envisioned the particle spectrum to
be composed of fireballs consistent of fireballs.  The idea was
that very heavy hadronic resonances decayed into somewhat less
heavy resonances, which would in turn decay to even lighter 
resonances. He showed this effect in his seminal paper
\cite{Hagedorn:346206} using resonances up to the $\Delta(1232)$
baryon that the hadronic resonances followed an exponential
mass spectrum, demonstrated by experiments over the course of
the last two decades as more and more particles have been
identified \cite{Broniowski:2000bj,Noronha-Hostler:2016ghw,ManLo:2016pgd,Zyla:2020zbs}.

To study the strong interaction further, heavy-ion collision
experiments collide atomic nuclei at relativistic speeds and
track their collective flow through charged particles.
Heavy-ion collisions offer a unique opportunity to
study an out-of-equilibrium many-body system, which crosses
the QCD phase transition from deconfined quarks and gluons
into hadrons. To accurately 
model a heavy-ion collision, several different stages
involving different physical phenomena have to be used:
initial condition, pre-equilibrium, hydrodynamics, hadronization,
and hadronic afterburner. Through out the modeling of heavy-ion
collisions, it is important that the equation of state is
consistent with the rest the modeling. At the point of
hadronization one must switch from quarks and gluons as
the degrees of freedom into hadrons.  Those hadrons and
their interactions must be consistent with the hadronic
part of the equation of state. This implies that if one
creates a state-of-the-art equation of state but uses a
hadronic afterburner with a mismatch in particles, it can
cause a number of problems. Thus, theorists are careful
to always match the hadrons in the equation of state (EOS)
to that of the hadronic afterburner (see \cite{Moreland:2015dvc,Alba:2017hhe,Karthein:2021nxe}).

%\jaki{Only use citations format for references, please change}
%\jordi{Updated all citations}.
%\jaki{this paragraph needs to be reformatted, you introduce SMASH but then switch gears entirely. There should be a separate SMASH paragraph and a different paragraph on the need for missing particles}
%\jordi{Added some more context to SMASH and some text to the next paragraph too}.
Simulating Many Accelerated Strongly-interacting Hadrons
(SMASH) is a state-of-the-art hadron transport code that
is widely used as afterburner after hydrodynamic simulations
and standalone for relatively low-energy heavy-ion collision
simulations \cite{Weil:2016zrk,dmytro_oliinychenko_2021_5796168}.
The approach followed by SMASH is based on
the relativistic Boltzmann equation, where the collision
term in the low-energy regime is dominated by binary hadron
scatterings and excitation and decay of resonances, i.e.,
by $2\leftrightarrow2$ and $1\leftrightarrow2$ reactions,
respectively, where the degrees of freedom are the
well-established hadronic resonances and their corresponding
properties \cite{Weil:2016zrk,Staudenmaier:2020xqr}. Recently,
SMASH has been used to investigate the effects of a high-density
medium on fluctuation observables, hadronic potentials, jet
quenching, and baryon, photon and deuteron production among
others, leading to similar results to those of other
implementations of microscopic
transport models \cite{Weil:2016zrk,Staudenmaier:2020xqr,
Mohs:2019iee,Schafer:2020vvw,Oliinychenko:2020znl,
Elfner:2020men,Sorensen:2020ygf,Staudenmaier:2021lrg,
Schafer:2021slz,Reichert:2021ljd,Schafer:2021csj,
Hammelmann:2022yso,Hirayama:2022rur}.

It has been shown that the inclusion of more hadronic resonances
when creating EOSs leads to significant changes in transport
coefficients \cite{Noronha-Hostler:2008kkf,Noronha-Hostler:2012ycm,Rais:2019chb,McLaughlin:2021dph}
and observables like the elliptic flow coefficient $v_2$ \cite{Noronha-Hostler:2013ria}, %\jaki{define v2}
susceptibilities \cite{Alba:2017mqu}, $p_T$ spectra 
\cite{Alba:2017hhe,Devetak:2019lsk}, and chemical freeze-out
conditions \cite{Alba:2020jir}, especially in the strange sector.
%\jaki{this seems like an introduction paragraph, it doesn't really fit here} %\jordi{Moved}
For example, in Refs.~\cite{Alba:2017mqu,Bazavov:2014xya}
the $\mu_S/\mu_B$ ratio was calculated to leading order
in $\mu_B$, as function of
susceptibilities of conserved charges, within the Hadron
Resonance Gas (HRG) model using different resonance lists,
showing a better agreement with lattice data up to the
transition temperature when more hadronic states are
considered. Moreover, the authors of \cite{Alba:2017mqu} also 
demonstrate that other related observables,
such as $\chi_4^S/\chi_2^S$ and $\chi_{11}^{us}$, indicate that
that the inclusion of addition of $|S|=1$ baryons and mesons 
is favorable, as opposed to multi-strange resonances.

Current equations of state use the PDG16+ particle list
\cite{Alba:2017mqu}, an exhaustive compendium of resonances
and their properties taken from the Particle Data Book
\cite{ParticleDataGroup:2016lqr}.
%\jaki{wrong citation, need the 2016 list}
With the motivation of using a transport code that includes
the same resonances as the EOS used for hydrodynamic stages,
we have revised the PDG16+ to create the PDG21+ list, which
has the updated masses and decays of all known experimentally
measured particles. In total, $24$ new particles
were added --mostly in the strange baryon sector-- and $10$ were
taken out, modifying the Hagedorn spectrum, as well as updating
previous- and adding several new decay channels. Massive
resonances in particular have been shown (experimentally) to
decay into three or four particle decays. 
%\jaki{explain SMASH can only do 1->2}
On the other hand, due to the geometrical
collision criterion used in SMASH, only $1\rightarrow2$ decays
are normally considered, with the exception of recent efforts
to implement a stochastic treatment \cite{Staudenmaier:2021lrg,
Garcia-Montero:2021haa}. The PDG21+ list was then adapted
to be used in SMASH using intermediate states to account
for multi-body decays.
%\jaki{Summarize major changes, number of new particles, number of particles that were changed etc}
%\jordi{Added a sentence a few lines above}

% In Sec.
% 2 we describe the development of the list and show a comparison
% to its predecessor, in Sec. 3 we show the impact of the inclusion
% of new resonances in SMASH, and in Sec. 4 we present our conclusions.

\section{PDG21+ resonance list}
%\jaki{this is misleading since this format has been around in the field much longer than the PDG16+.  Instead, just focus on what you did and what is needed: The development of the new PDG21+ list was possible because of the existing PDG16+ precedent,} 
%\jordi{My intention here was to mention how we have more information now and an update was needed. Then I proceed to say how we carried out the update and what was changed. How can I convey this message in a better way?}
Over the past 5 years experiments have shed light on
new particle resonances, providing better information on their
masses and known decay channels compared to what was known in 2016
when the PDG16+ was created. Here we build on the previous PDG16+
list that includes the particles and properties, including particle
ID (PID), mass, width, degeneracy, baryon number, strangeness content,
isospin, electric charge, and branching ratios of decay channels.
The PDG16+ contemplated $408$ different particles, of which
$153$ were mesons and $255$ are baryons. 
%\jaki{I'm confused what is the $N$ supposed to be here?s}
%\jordi{I left them as placeholders but did not have the time to fill them in before}

An extensive revision of the PDG16+ was carried out, updating
the values of mass and width, as well as decay channels and
branching rations to the most recent experimental data available.
For heavier resonances (mass $\gtrsim 1.5$ GeV), it becomes more
and more common to have missing decay channels, i.e., the reported
branching ratios do not add up to 1. In the 16+ edition of the list,
a ratio of $90\%$–$10\%$ was assigned for unknown and known decay
channels, respectively, where unknown decay channels were modeled
as radiative decays to a relatively lighter hadron. In the case of
the 21+ edition, the experimentally reported ratio was kept as is,
only using radiative decays as a complement to obtain the $100\%$
of decays. Recent experimental results, such as the
observations in \cite{Sarantsev:2019xxm, Hunt:2018wqz, 
CBELSATAPS:2020cwk}, provided new knowledge of the branching
ratios of heavy resonances, especially in the $\Sigma$ and $\Lambda$
sectors, thus relaxing the need of approximations.
%\jaki{make clear if this was needed or not. If I remember correctly most of the branching ratios were known so we didn't need to do this anymore?}
%\jordi{Added a sentence but I'm not sure if it addresses your concern completely. It was necessary to introduce some kind of modeling, yes, but we now thought that keeping the ratios was less invasive than fixing a 90-10 branching ratio hierarchy.}

%\jaki{Explain star scheme starting with 4 stars etc.  Then explain why you need 1-2 star states using "As was shown in \cite{Alba:2017mqu}," only after that explain the below}
In the Review of Particle Physics, particles are
organized according to a confidence level scale, depending
on the amount of evidence to back up the existence of each
particle and their properties. The most well-established
states are marked with four stars (****), whilst resonances that
have minimal information are given one star (*). As was shown in 
\cite{Alba:2017mqu}, 1-2 star states are fundamental to 
describe lattice results. To qualify as an
entry for the PDG21+, it was generally sufficient for the
candidate resonance to have one star of confidence level and to be located under the Particle Listings section of the Particle Data Book \cite{Zyla:2020zbs}. Notice that some particles in the Review
are labeled as \textit{Further States} and are not included in
the PDG21+ due to the overall lack of information for such states. 
Moreover, some states in the Listings section have also been omitted;
such is the case of $\Lambda(2585)$ or $\Sigma(3000)$. Only light and strange hadrons are
considered for particle and decay channel listings, leaving charm
and bottom hadrons out, as well as leptons. The new version of the
list contains $418$ different particles, of which $151$ are mesons
and $267$ are baryons. 
%\jaki{Do you mean to fill in the N's?}
%\jordi{Yes}

\begin{table}[h]
 \centering

 \begin{tabular}{|c|c|}
 \hline
 Particle name & Status  \\ \hline\hline
 $a_1(1420)$ & deleted   \\ \hline
 $X(1840)$ & deleted       \\ \hline
 $\pi_2(2005)$ & added  \\ \hline
 $X(2370)$ & added      \\ \hline
 $a_6(2450)$ & deleted  \\ \hline
 $\Sigma(1770)$ & deleted  \\ \hline
$\Sigma(1840)$ & deleted  \\ \hline
$\Sigma(2000)$ & deleted  \\ \hline
 $\Sigma(2010)$ & added \\ \hline
 $\Sigma(2160)$ & added  \\ \hline
 $\Sigma(2230)$ & added  \\ \hline
 $\Sigma(2455)$ & added \\ \hline
 $\Sigma(2620)$ & added \\ \hline
 $\Sigma(3170)$ & added \\ \hline
 $\Lambda(2070)$ & added \\ \hline
 $\Lambda(2080)$ & added\\ \hline
$\Omega(2012)$ & added \\ \hline
 \end{tabular}
 \caption{Newly added and deleted particles in the PDG21+ list. 
 It is understood
 that each particle includes all the elements of the corresponding
 multiplet and their antiparticles.}
 \label{tab:particles_added}
\end{table}

\begin{table}[h]
\centering
\begin{tabular}{|c|}
\hline
Particle name   \\ \hline\hline
$\Sigma(1730)\rightarrow\Sigma(1780)$   \\ \hline
$\Sigma_+(1940)\rightarrow\Sigma(1940)$       \\ \hline
$\Sigma_-(1940)\rightarrow\Sigma(1910)$  \\ \hline
$\Lambda(2020)\rightarrow\Lambda(2085)$  \\ \hline
\end{tabular}
\caption{Renamed particles in the PDG21+ list. It is understood that each particle
includes all the elements of the corresponding multiplet and their antiparticles.}
\label{tab:particles_renamed}
\end{table}

\begin{figure}[h]
 \includegraphics[width=\linewidth]
    {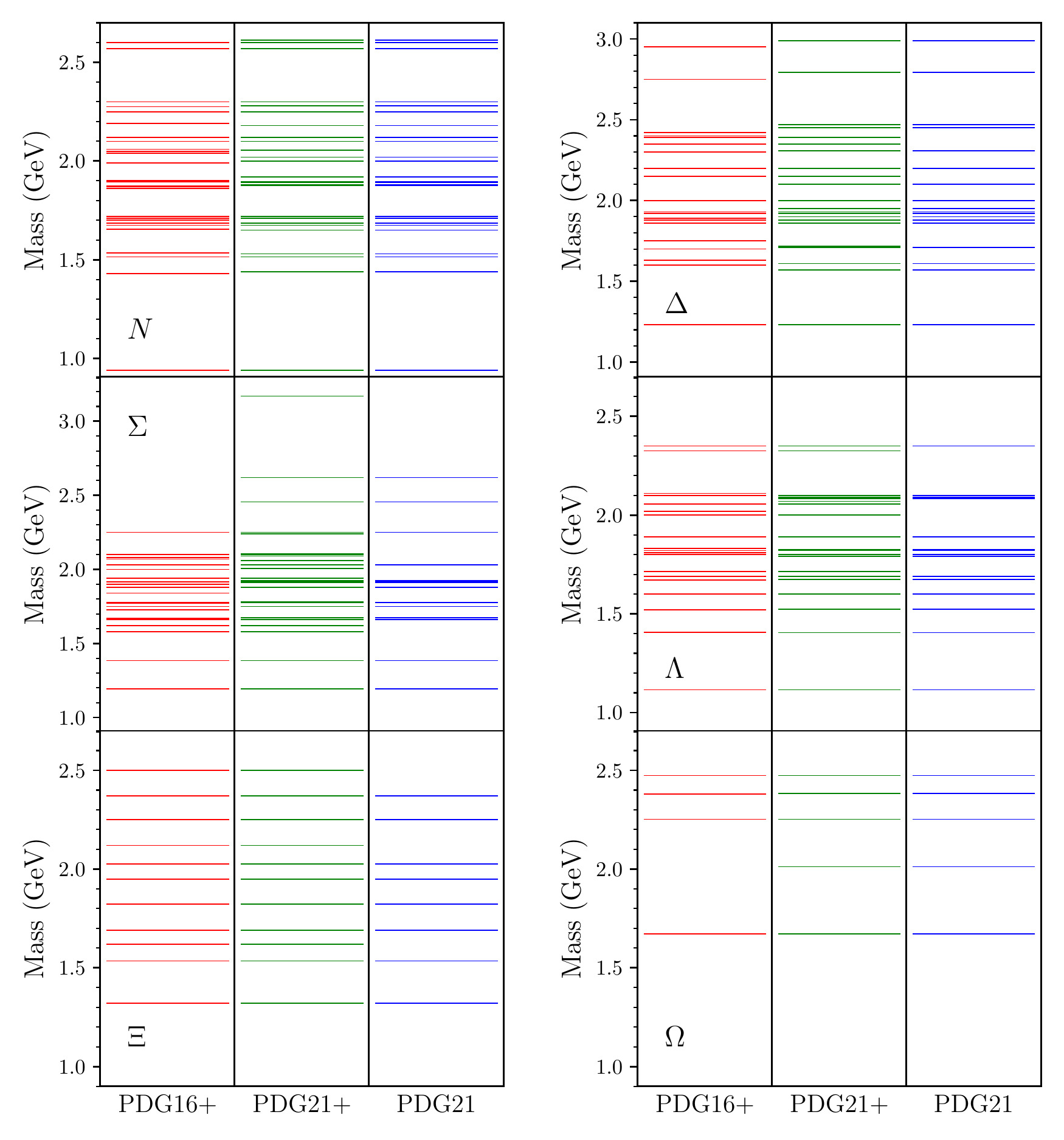}
 \caption{Particle spectra per species as extracted from
 the PDG16+ and PDG21+ lists. The newer list contains more
 resonances and updated properties for previously known particles.}
 \label{fig:PDG_spectrum}
\end{figure}

%\jaki{For table I and II you can merger into one table and just label as particles added and deleted}
%\jordi{Ok, done}
A total of $24$ particles were added and $10$ were removed with 
respect to the PDG16+ list. In Table~\ref{tab:particles_added}, a
complete list of the particles added and removed are shown. 
In addition, some particles were renamed, and are shown in 
Table~\ref{tab:particles_renamed}; these are states that
kept the same quantum numbers but had their mass updated
under newly available experimental information. In 
Fig.~\ref{fig:PDG_spectrum} we present a comparison of the
particle spectra per hadronic species between the previous
PDG16+ and the new PDG21+ lists, including the more restrictive
PDG21 version, which only includes states with a 3-star degree
of certainty or more; it is clear that more resonances have been
included and of particular interest, particles with strange
content, which are precisely the ones where lattice results
suggested new resonances could better explain the data.

%\jaki{show examples of some decay channels}
%\jordi{Is this the best place to do so? I have one in the following section. Should I add more?}
%\jaki{include exponential mass spectrum, explain how that works}
%\jordi{Added a couple of paragraphs, please check}
Hadron resonance gas models 
are commonly used to study the thermodynamics of the hadronic
phase of QCD matter, which can be extracted from the a partition
function whose only free parameters are the number of
particle resonances and their masses. Early on, it was predicted by Hagedorn that
the number of hadronic resonances with respect of mass would grow
exponentially, that is,
\begin{equation}
\label{eq:Hagedorn_spectrum_general}
 \rho(m) = f(m)\exp(m/T_H),   
\end{equation}
where $f(m)$ is a slowly varying function of $m$ and $T_H$ is 
a free-parameter known as the Hagedorn limiting temperature, 
understood soon after as the temperature at which the hadronic 
description breaks down \cite{Broniowski:2000bj,Noronha-Hostler:2016ghw,ManLo:2016pgd}. 

\begin{figure}[h]
 \includegraphics[width=\linewidth]
    {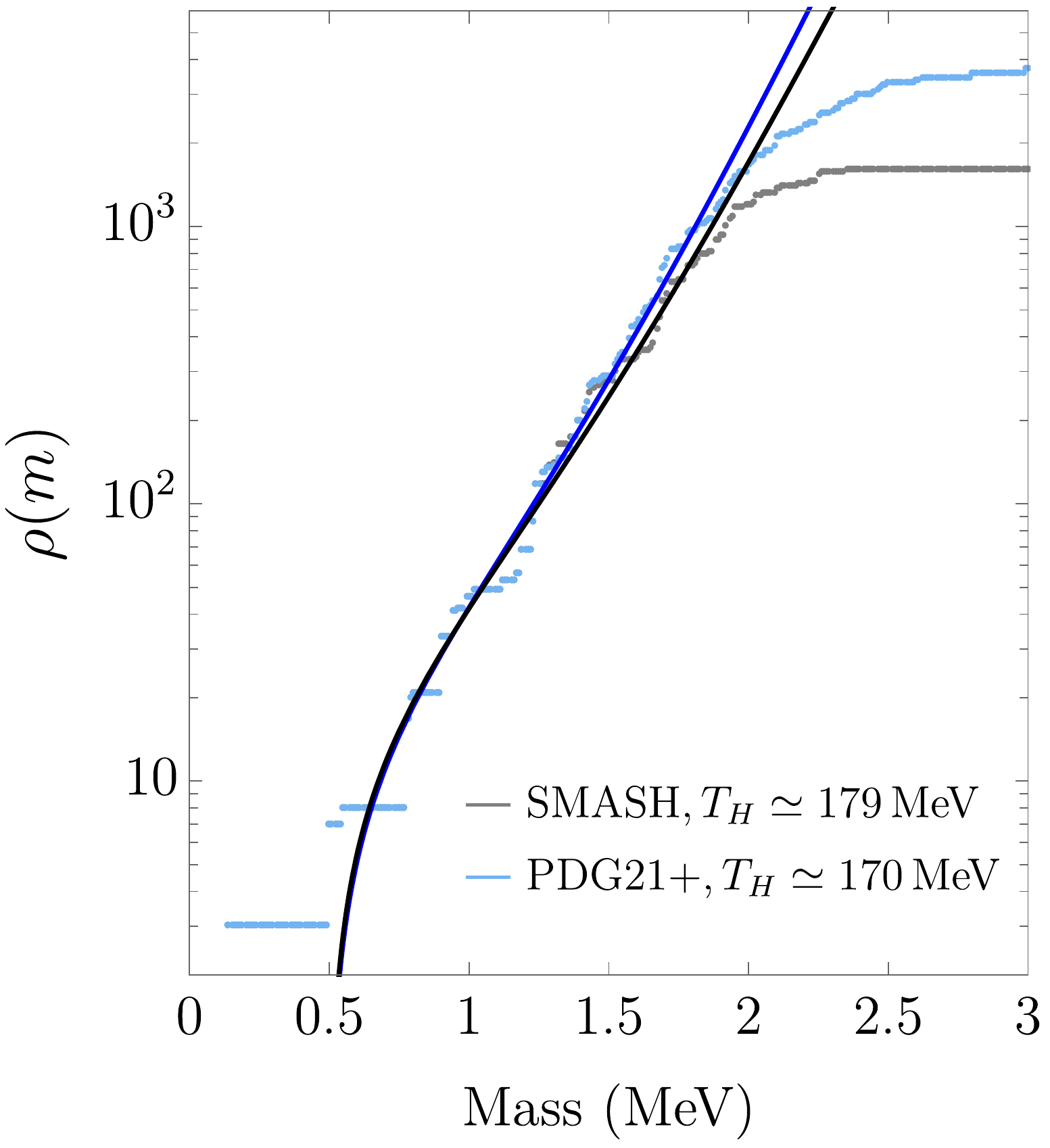}
 \caption{Particle spectra for the well-established
 states included in SMASH (gray) and the full PDG21+
 list (light blue) with the corresponding
 fits of Eq.~\ref{eq:Hagedorn_spectrum} shown in black and
 blue, respectively, and the respective Hagedorn temperatures.}
 \label{fig:hagedorn_spectrum}
\end{figure}

Despite its simplicity, the HRG
model is in good agreement with lattice data up to temperatures
close to the phase transition temperature \cite{Noronha-Hostler:2013ria} and
presents itself as a tool to create new EOSs. Hence, it becomes
necessary to study the behavior of the particle spectrum 
as new particles are considered, since these changes will have
an effect on the corresponding Hagedorn temperature. In 
Fig.~\ref{fig:hagedorn_spectrum} we show the particle spectra
for all particles in the PDG21+ list, compared to those included
on the default SMASH release, along with the corresponding
fits to data, using a version of Eq.~\ref{eq:Hagedorn_spectrum_general}
given by
\begin{equation}\label{eq:Hagedorn_spectrum}
    \rho(m) = \int_{m_0}^m \frac{A}{\left[M^2+M_0^2 \right]^{5/4}}\,e^{M/T_H}dM
\end{equation}
Although the extracted limiting temperature is highly sensitive
to the mass cutoff, it is found that the Hagedorn temperature
is lower in the case of the new list, $T_H^\text{PDG21+}
\simeq170$ MeV, than the one coming from the particle set in SMASH, $T_H^\text{SMASH}\simeq179$ MeV, approaching the phase
transition temperature of $\sim 155$ MeV.

%\jaki{Table \ref{tab:particles_renamed} has not been discussed, please do so}
%\jordi{It was only mentioned previously but not discussed. Added an additional sentence and moved the table closer to where it is referenced in the text}

\section{Implementation in SMASH}
The Quark Gluon Plasma is described by relativistic viscous hydrodynamics that rapidly expands and cools in time until particles cool enough to become hadrons.  After hadronization, the particles are still interacting with other hadrons, which can be described through hadron transport models. These models are crucial to make direct comparisons to experimental data because experiments can only measure the final state particles. 

Here we use the hadron transport code, SMASH \cite{Weil:2016zrk,dmytro_oliinychenko_2021_5796168} that is an open-source code written in C++ that is commonly used in the field following the hydrodynamic phase \cite{Ryu:2019atv,Oliinychenko:2018ugs,JETSCAPE:2020mzn,Schafer:2021csj,Wu:2021fjf,Schafer:2021slz}.
The current version of SMASH has $222$ states and their decay channels. Because both PDG16+ and PDG21+ have significantly more states, it is not possible to run an EOS, such as \cite{Parotto:2018pwx,Noronha-Hostler:2019ayj}, that is based upon the PDG16+ list in a hydrodynamic model and then use SMASH in its current format. 

The reason for the mismatch between the most current PDG  lists and SMASH is because it is not straightforward to simply add new, heavy resonances into SMASH.  The main barrier is that SMASH cannot handle $1\rightarrow 3$ or $1\rightarrow 4$ body decays. There are computational difficulties in handling the back reactions, i.e., $3\rightarrow 1$ and $4\rightarrow 1$ and without those back reactions detailed balanced would no longer be preserved. However, eliminating particles with these channels also isn't the solution since many heavy resonances decay into 3+ particles. Thus, the solution has been to convert $1\rightarrow 3$ or $1\rightarrow 4$ into decays with an intermediate resonance that then provides the same final state. For example,  the $f_0(1500)$ hadron has 12 decay channels and
one of them goes to $\pi^+\pi^+\pi^-\pi^-$. We can then include an intermediate decay $f_0(1500) \rightarrow \rho^0\rho^0$, since each $\rho^0$ meson will then also decay into two pions, i.e., $\rho^0\rightarrow \pi^+ \pi^-$. In this manner the final state has been preserved, although the decay itself is slightly slowed down by passing through an intermediate channel. 

%\jaki{you need to mention something about angular momentum choices}
%\jordi{Added a sentence in the following paragraph}

%\jordi{Should we delete the first sentence of this paragraph? I feel it is redundant now that the previous paragraph talks about intermediate states} 
%In order to continue in our goal to be consistent with EOSs and to overcome the lack of 3 and 4-body decays, we have modeled these multi-particle decays as a sequence of $1\rightarrow2$ decays, with an intermediate state that could be used in SMASH.
The correct identification of an intermediate state
required a number of steps. For each heavy resonance
with a 3 or 4-body decay, the least massive 2-particle intermediate state that could
further decay into the final state was chosen. 
Furthermore, the lowest possible absolute value
for the angular momentum $L$ was also chosen for each
decay channel, depending on the daughter particles, since
SMASH enforces angular momentum conservation in decays. In the few cases
where was no possible intermediate state, or it violated mass
conservation, the mass of the parent particle
was increased as long as the increment was small compared
to the original mass. Nonetheless, wherever this was not
viable, such decay channels were deleted, having all other
branching ratios normalized; this was the case for
25 decay channels. 
%\jaki{I thought sometimes you also increased the original particle's mass? How many total were deleted?} 
%\jordi{In most cases we were able to include the intermediate state with no additional modification; in some other cases we needed to bump up the mass and were ok, and in a smaller subset, there was simply no way of increasing the mass without messing up other things} 
In cases where many possible intermediate
states could in principle decay to the final state, only the
one with the lowest combined mass was chosen, as it is the most
energetically favored state.

One challenge with this method is that intermediate states may
have other possible decay channels besides the preferred one.
For example, the hyperon resonance $\Lambda(1690)$ decays
into $\Sigma+\pi+\pi$ about $20\%$ of the time, whose
intermediate state is $\Sigma(1385)+\pi$, using $\Sigma(1385)$
as a proxy for the $\Sigma+\pi$ pair. However, the $\Lambda+\pi$
decay channel for the $\Sigma(1385)$ resonance is more usual
than $\Sigma+\pi$.
%\jaki{Jordi give an example here}

To handle the connection between the PDG, standard formats used
in the heavy-ion community, and SMASH, we have written a code
that converts the output of our original table of particles
(taken diretctly from the PDG) into the correct format, making
all the adjustments discussed here.  This code can output formats
compatible with ThermalFIST \cite{Vovchenko:2019pjl} such that
the PDG21+ can be used in thermal models as well. This new code
also allows one to easily add new particles from further upgrades
to the PDG such that we do not anticipate that one should have
to wait every 5 years to upgrades the particle lists used within
the heavy-ion community.

Once the new particles are in the SMASH format, they can then
be implemented directly into SMASH.  The first thing to test is
the effect of these new states on the cross-sections.

\begin{figure}[H]
 \includegraphics[width=\linewidth]
    {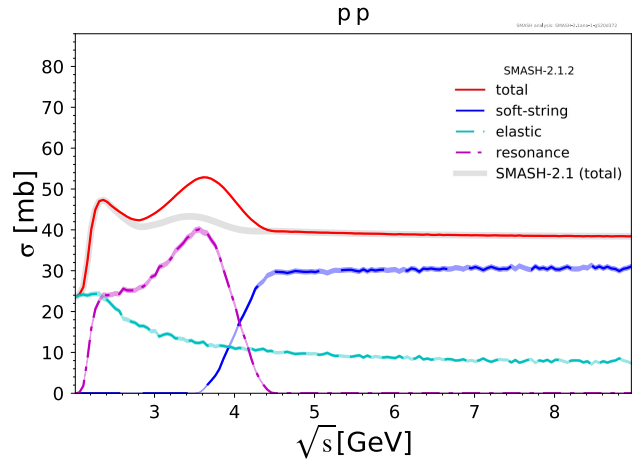}
 \caption{Proton-proton cross-section by type of process. The
 inclusion of new resonances and decay channels contribute
 to a bump in the total cross-section calculated using the PDG21+
 decay list (red line) compared to the default SMASH list (gray line).}
 \label{fig:pp_xs}
\end{figure}

\begin{figure}[h]
 \includegraphics[width=\linewidth]
    {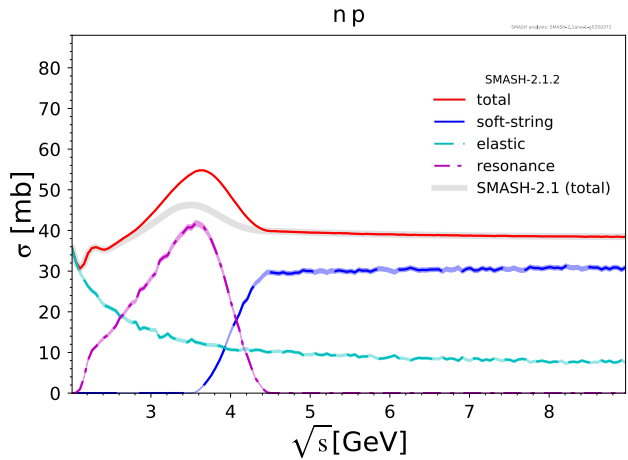}
 \caption{Neutron-proton cross-section by type of process. The
 inclusion of new resonances and decay channels contribute
 to a bump in the total cross-section calculated using the PDG21+
 decay list (red line) compared to the default SMASH list (gray line).}
 \label{fig:np_xs}
\end{figure}

The cross-sections have already been measured experimentally and,
therefore, they must reproduce experimental data. In
Figs.~\ref{fig:pp_xs} and \ref{fig:np_xs},
we present two examples, the $pp$ and $np$, cross-sections. The
red line is the total cross-section that has to match experimental
data, the cyan line is the contribution of elastic processes, magenta
stands for the contribution of resonances, blue is the contribution
of string processes handled with the help of Pythia \cite{SJOSTRAND2008852}, and
the gray line is the total cross-section without any of the 3 and 4-body
decays included in SMASH, which matches experimental data. 
The contributions coming from resonances
have increased significantly.  There is a distinct
bump in each of the plots, which occurs because of the inclusion of these new particles and their corresponding decay channels. Now that the effect of including more resonances in
SMASH has been shown, it is necessary to reproduce the experimental
data for the elementary total cross-sections, which requires adjustments on their treatment within SMASH.

\section{Conclusions}
Experimental developments  have lead to the
discovery of new heavy hadronic resonances and better knowledge of
their interactions over the past 5 years. In this work we have
updated the particle list to the PDG21+ that includes formats
that are compatible with both SMASH and ThermalFIST. As more
particles are taken into consideration when
building a new EOS, observables such as 
susceptibilities are modified, in many cases approaching a
better description of lattice data. However, to be fully 
consistent, one has to be careful and use the same particle
list and decays –as the one coming from the EOS– when using
afterburners. In particular, we have compiled a new particle
list, the PDG21+, with the latest Particle Data Book information
available and adapted it to work with SMASH. The latter was 
done by modeling 3 and 4-body decays as a sequence of 2-body
decays with intermediate states. To quantify the effect of 
new heavy resonances, we computed the total $pp$ and $np$
cross-sections, observing a bump coming from the newly added
channels and hadronic states. In order to adapt the list in
SMASH consistently with experimental data for elementary total cross-sections, more work is needed adjusting the internal framework; this
is currently underway and the results will be
published elsewhere. We will then explore the consequences of the addition of these new resonances both with SMASH comparisons to experimental data at low beam energies as well as hydrodynamics coupled to SMASH using a hybrid approach at high energies. 

\section{Acknowledgments}
This work was supported in part by the NSF within the framework of the MUSES collaboration, under grant number OAC-2103680.
J.N.H. acknowledges the support from the US-DOE Nuclear Science Grant No. DE-SC0020633. The authors also acknowledge support from the Illinois Campus Cluster, a computing resource that is operated by the Illinois Campus Cluster Program (ICCP) in conjunction with the National Center for Supercomputing Applications (NCSA), and which is supported by funds from the University of Illinois at Urbana-Champaign. 
J.H. acknowledges the support by the DFG SinoGerman project (project number 410922684).

\nocite{*}
\bibliography{arxiv_version.bbl}

\end{document}